\documentclass[aps,amsmath,amssymb,prl,twocolumn,showpacs,superscriptaddress]{revtex4}
\usepackage[latin1]{inputenc}
\usepackage{graphicx}
\usepackage{bm}

\begin{document}

\title{A model of hyphal tip growth involving microtubule-based transport}

\author{K. E. P. Sugden}
\affiliation{SUPA, School of Physics, University of Edinburgh, Edinburgh EH9 3JZ}
\author{M. R. Evans}
\affiliation{SUPA, School of Physics, University of Edinburgh,
Edinburgh EH9 3JZ}
\affiliation{Isaac Newton Institute for Mathematical Sciences,
Cambridge, CB3 0EH}
\author{W. C. K. Poon}
\affiliation{SUPA, School of Physics, University of Edinburgh,
Edinburgh EH9 3JZ}
\author{N. D. Read}
\affiliation{Institute of Cell Biology, University of Edinburgh,
Edinburgh EH9 3JH}

\begin{abstract}
We propose a simple model for mass transport within a fungal hypha and
its subsequent growth.  Inspired by the role of
microtubule-transported vesicles, we embody the internal dynamics of
mass inside a hypha with mutually excluding
particles progressing stochastically along a growing one-dimensional
lattice. The connection between long range transport of materials
for growth, and the resulting extension of the hyphal tip has not
previously been addressed in the modelling literature.  We derive and
analyse mean-field equations for the model and present a phase diagram
of its steady state behaviour, which we compare to simulations.  We
discuss our results in the context of the filamentous fungus, {\it
Neurospora crassa}.

\end{abstract}

\date{\today}

\maketitle

Biologically, fungi are distinct from both plants and animals.  In
addition to their intrinsic interest, they impact immensely on human affairs
and on the ecosystem \cite{deacon}.

Key to the evolutionary success of fungi, is their unique mode of
growth.  Filamentous fungi grow by the polarised extension of
thread-like hyphae, which make up the body, or mycelium, of a
fungus. Except for branching (which initiates new hyphae) the site of
growth is localised to a single region at the tip of each elongating
hypha. 

There are many theoretical models for the growth of fungal colonies
and of single hyphae (reviewed in \cite{Prosser,bezzi}). Most models
of single hypha growth concentrate on bio-mechanics \cite{Koch,GT}. Of
more interest for us here is the ``vesicle supply centre'' (VSC) model
\cite{Garcia,TKM}, in which raw materials for growth are packaged in
secretory vesicles and distributed to the hyphal surface from a single
``supply centre'' (often identified with an organelle complex known
as the Spitzenk\"{o}rper, or apical body \cite{26,27}) situated within the
growing tip. This model is capable of predicting the
shape of hyphal tips; but the speed of growth (equivalent to the speed
of the VSC) is an input parameter. Moreover, all transport processes
are subsumed into a single rate of vesicle supply at the VSC. We are
aware of just one model that takes explicit account of transport along
the growing hypha \cite{Trinci79}. A major interest of this early
work, however, was the initiation of branching; these
authors did not relate vesicle transport to growth velocity. This
latter issue remains poorly understood.

In this work, we propose a simple one-dimensional model which makes
an explicit connection between the long-distance transport of building
materials along a hypha and the resulting extension as they are
delivered to its apical site of growth.

It is a highly idealised model, encompassing many complicated
biological processes (many of which are still poorly understood) with
two key parameters: the rate at which vesicles enter the system and
the efficiency with which they extend the length of the hypha. We
demonstrate that by altering these rates, steady states can be
attained whereby the hypha is extending at a constant speed while
being supplied with materials far behind the tip.  Our model has
features in common with \cite{Trinci79}. Like \cite{Trinci79}, we use
computer simulations, but unlike \cite{Trinci79}, we can also make
analytical progress because of recent advances in non-equilibrium
statistical physics.

Our model is inspired by the well-known Totally Asymmetric Simple
Exclusion Process (TASEP).  This is a one-dimensional lattice along
which particles progress through stochastic directed motion.  No more
than one particle may occupy each lattice site at any given time.  The
TASEP was originally introduced as a lattice model of ribosome motion
along mRNA \cite{MGP} and recent variants have been widely used to
model the collective dynamics of molecular motors
\cite{KL,PFF,KKCJ,Yariv}. The application of this and other classes of
statistical mechanical models to many kinds of ``biological traffic''
has recently been reviewed \cite{Chowdhury}.  The TASEP is also widely
studied in its own right as a fundamental model of non-equilibrium
statistical mechanics \cite{Mukamel}, in particular as a simple driven
diffusive system exhibiting non-equilibrium phase transitions
\cite{Krug} between different macroscopic density and current regimes
\cite{DDM,TASEP}. Our work contributes to the study of both
``biological traffic'' and non-equilibrium phase transitions.

In constructing the model, we introduce a new feature into the TASEP:
particles reaching the end of the lattice act to extend it.  We ask
whether a constant input rate far from the growing end of the lattice
can generate steady state lattice growth and if so, how the growth
velocity depends upon the system parameters.  We find that, as in the
TASEP, different macroscopic regimes exist in the growing system, with
different forms for the growth velocity, and non-equilibrium phase
transitions between these regimes. We ask whether these steady states
may be related to the growth states observed in fungal hyphae.

\begin{figure}
\centering
\includegraphics[scale=0.2]{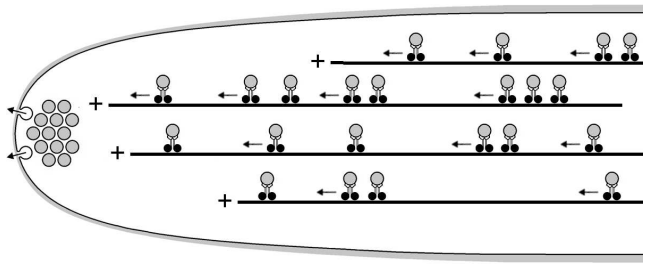}
\caption{\label{hyp}A simplified diagram of the process of polarized
  secretion resulting in tip growth in a fungal hypha.  Secretory
  vesicles are transported by kinesin molecular motors (not to scale)
  along microtubules towards their growing ``plus'' ends.  The vesicles
  accumulate within the so-called Spitzenko\"rper at the hyphal tip
  before fusing with the apical plasma membrane.  The secretory
  vesicles deliver membrane proteins and lipids, cell wall
  synthesizing enzymes and possibly cell wall precursors to the
  growing fungal tip.  Hyphal widths typically vary between 5 and 15~$\mu$m.}
\end{figure}

To arrive at the model we appeal to the popular belief that the long
distance transport mechanism within a hypha is provided by kinesin
molecular motors, ``walking'' along microtubule filaments which run
length-ways through the hypha \cite{Seiler}.  Once at the tip,
vesicles fuse with the plasma membrane, resulting in localised
extension of the hyphal wall \cite{Gow,26,27}.  Fig.~\ref{hyp}
summarises this process diagrammatically.

A kinesin motor with cargo progressing toward the tip will likely
attach and detach between a number of microtubules on its way, and is
then thought to be transferred onto actin filaments at the VSC, before
being distributed to the hyphal tip \cite{26}. In our simple model
however, we shall bypass the details of the vesicle trajectory and
concentrate instead on a coarse-grained description representing what
is essentially a continuous directed movement of vesicles toward the
hyphal tip.  We thus assume here simply that a continuous driving
force is applied to the vesicles, all the way to the point of fusion.

We construct the model by supposing that a hypha contains a number of
\emph{effective} microtubule tracks which run \emph{continuously} to
the tip.  Each effective microtubule is modelled by a 1-D lattice,
with the motors plus cargo represented by particles which progress
along the lattice sites.  We identify lattice site $1$ with the hyphal
tip. Particles obey hard-core exclusion and move in one direction
only, toward the tip, without overtaking.  When particles leave lattice site
$1$, they act to extend the lattice through the transformation:
\emph{particle $\rightarrow$ new site}.

We justify our model with a simple order of magnitude test. We
identify the lattice repeat with the kinesin step size, 8~nm
\cite{Bray}. In the model organism {\it N. crassa}, a
10~$\mu$m wide hypha growing at $25^\circ$~C has as extension rate of
$\sim0.5$~$\mu$m per second.  It has been estimated that $\sim600$
vesicles per second must fuse with the tip to provide enough plasma
membrane and other materials to maintain this hyphal extension rate
\cite{Trinci74}. Thus, the extension of a hypha by one model lattice
unit ($\sim 10$~nm) is equivalent to the arrival of order 10
vesicles. If each particle delivered to the end contributes to lattice
extension, we require $\sim 10$ equivalent effective microtubules in a
typical hyphal cross section.  This is acceptably within an order of
magnitude of the number of microtubules observed near the
Spitzenko\"rper within the growing hyphal tip of {\it N. crassa} \cite{microtubules}.

\begin{figure}
\centering
\includegraphics[scale=0.7]{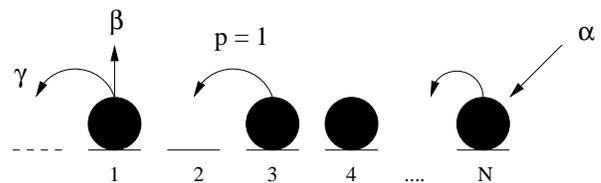}
\caption{\label{model} Schematic of the model with input rate
$\alpha$, hopping rate $p=1$, absorption rate $\beta$ and growth rate
$\gamma$.}
\end{figure}

The model dynamics are specified by the rates at which the following
processes occur on the lattice: particles in the bulk hop toward the
tip with rate 1; particles enter the lattice far from the tip with
rate $\alpha$; particles detach from site $1$ with rate $\beta$ and
transform into a new lattice site with rate $\gamma$, as shown
schematically in Fig.~\ref{model}.  Thus $\gamma$ is the parameter
controlling the lattice growth and $\beta$ allows particles to leave
the end of the lattice without extending it.  Ratio $\gamma/\beta$ is
thus the efficiency with which the hypha extends.  Biologically,
$\gamma$ represents the rate at which secretory vesicles fuse with the
hyphal tip, $\alpha$ represents the vesicle density far from the tip
and $\beta$ allows vesicles to reach the tip without contributing
directly to growth.

\begin{figure}
\centerline{
\mbox{
\includegraphics[width=2.5in,angle=270]{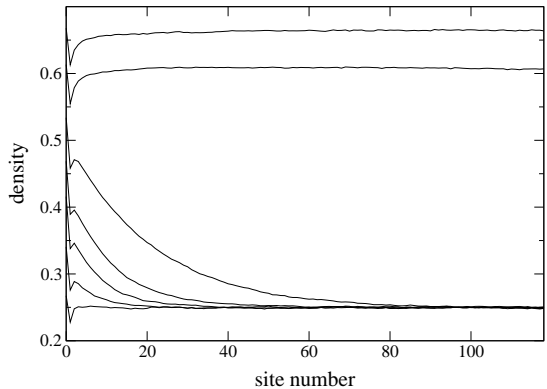}
}
}
\caption{Average site occupancy for $\alpha = 0.25$. Upper two traces 
are for $\gamma = 0.2$ (higher) and 0.24 (lower). 
Lower five traces (highest to lowest) for $\gamma = 0.28-0.56$ display 
apical peaks and decays.}
\label{MC} 
\end{figure}

We initially perform Monte Carlo (MC) model simulations by
stochastically updating particles on a single lattice according to the
above dynamics. After some relaxation time, density profiles are
obtained by averaging site occupancies over many updates.  We find
three different macroscopic behaviours.  Results for representative
parameter values $\alpha = 0.25$, $\beta=0$ and $\gamma$ in the range
0.2 to 0.56, are shown in Fig.~\ref{MC}.  For high values of $\gamma$
one sees profiles that decay from the tip to a $\gamma$-independent
bulk density equal to $\alpha$.  For the highest values of $\gamma$
the density at the tip is $<\alpha$.  As $\gamma$ is lowered the tip
density is $>\alpha$ and the region over which the decay occurs grows
in size. For low values of $\gamma$ we see distinct profiles where the
bulk density is $\gamma$-dependent and is $\gtrsim 0.4$ (these profiles were
also seen for low $\gamma$, high $\alpha$).  The transition from the
high $\gamma$ to the low $\gamma$ profiles is discontinuous and
involves a jump in the bulk density.  In the regime of high $\alpha$
and $\gamma$ (not shown), density profiles with algebraic decays
between the boundaries were observed.  These are ``maximal current''
profiles, which we shall discuss shortly.

We now seek an analytical understanding of our observations using a
mean-field approximation where we consider the average density,
$\rho_i(t)$ at site $i$, and ignore correlations between the density
at different sites \cite{DDM}.  We describe the growth dynamics in a
frame of reference co-moving with the growing tip.  The tip site is
always labelled site $1$.  Each time growth occurs, all other site
labels must therefore be updated $i\ \rightarrow\ i+1$.  The change in
density at site $i$ is the net result of particles entering from the
site on the right, departing to the site on the left and shifting 
right due to index relabelling during growth.  Within the mean-field
approximation, we have, for $i>2$,
\begin{equation}
\label{rhoi}\frac{d\rho_i}{d t}
=\rho_{i+1}[1-\rho_i]-\rho_i[1-\rho_{i-1}]+\gamma\rho_1[\rho_{i-1}-\rho_i]\;.
\end{equation}

Note that the third term is proportional to the rate at which lattice
sites are added, $\gamma\rho_1=v$ which is the tip velocity.  Separate
equations govern the change in density at sites $1$ and $2$, in order
to take into account the effect of the \emph{particle $\rightarrow$
new site} transition:
\begin{eqnarray}
\label{Eq:rho1}\frac{d\rho_1}{d 
t}&=&\rho_2[1-\rho_1]-(\gamma+\beta)\rho_1 \;,\\
\label{Eq:rho2}\frac{d\rho_2}{d 
t}&=&\rho_3[1-\rho_2]-\rho_2[1-\rho_1]-\gamma\rho_1\rho_2\;.
\end{eqnarray}

Eq. \ref{Eq:rho2} differs from the bulk equation only in that should a
growth event occur, there is never particle at site $2$ after the
lattice indices are updated.  The final term is hence a decrease in
density at site $2$.

Finally, since particles enter at rate $\alpha$, the particle density
at the right-hand end is effectively $\alpha$. As the lattice grows,
this boundary recedes from the tip with velocity $-v$, ultimately
corresponding
to boundary condition
\begin{equation}
\lim_{N\to \infty} \rho_N = \alpha\;.
\label{lbc}
\end{equation}

We seek a steady state solution for this system defined in the
reference frame of the tip.  Such a solution is characterised by a
constant current of particles everywhere through the system, a uniform
tip velocity and a density profile which decays to the right boundary
condition over a finite length scale, so that the profile is
effectively independent of the system size, i.e. we seek solutions to
(\ref{rhoi}--\ref{Eq:rho2}) with the time derivatives set to zero and
obeying the boundary condition (\ref{lbc}).  We obtain an expression
for the particle current through the system in the tip's stationary
frame from Eq. \ref{rhoi}
\begin{equation}\label{J}J=\rho_{i}[1-\rho_{i-1}]-v\rho_{i-1}\;.
\end{equation}

Now, at the tip $J= (\gamma+\beta)\rho_1 = (1+\beta/\gamma) v$,
whereas far away from the tip (\ref{J}) yields $J=
\alpha(1-\alpha-v)$, so that
\begin{equation}
v= \frac{\alpha(1-\alpha)}{1+\alpha + \beta/\gamma}\;,
\label{Eq:v2}
\end{equation}
which gives the tip velocity in terms of $\alpha$
and $\beta/\gamma$.

We now restrict ourselves to $\beta=0$, and comment on the effects of
non-zero $\beta$ later.  Since for $\beta=0$ the tip velocity is
simply a result of a flux of particles through the final lattice site,
we have $J=v$ and thus from (\ref{J}) a recurrence relation between
the steady state density at any site and that at the previous site:
\begin{equation}\label{RR}{\rho_{i}=\frac{v(1+\rho_{i-1})}{1-\rho_{i-1}}}
\quad i>2\;.
\end{equation}
We define $\rho_\infty=\alpha$ as the stable fixed point value to which
this
recurrence relation converges:
\begin{equation}
\label{alpha}
\rho_{\infty}=\alpha=\frac{1-v-\sqrt{1-6v+v^2}}{2}\;.
\end{equation}
The decay length to $\alpha$ is finite and independent of lattice
size, as required.  We are now able to solve for all densities in
terms of the parameters, $\alpha$ and $\gamma$:
\begin{eqnarray}
\rho_1=\frac{v}{\gamma}&=&\frac{\alpha[1-\alpha]}{\gamma\alpha+\gamma}\;,
\\
\label{Eq:rho22}\rho_2=\frac{v}{1-\rho_1}&=&\frac{\gamma\alpha[1-\alpha]}{\gamma[\alpha+1]-\alpha[1-\alpha]}\;,
\end{eqnarray}
and for $i>2$, $\rho_i$ is given through (\ref{RR}).

Under our constraint that the bulk density is $\alpha$, we find two
types of steady-state solution to the mean-field equations.  In these
solutions the profiles decay exponentially toward $\rho=\alpha$ and are
distinguished by whether $\rho$ decays to $\alpha$ from above or
below.  For $\gamma>(1-\alpha)/2$, which we refer to as region I, the
decay is from above and for $\gamma< (1-\alpha)/2$, which we refer to
as region II, the decay is from below, from a minimum value at site 2,
although there is a peak in the density at site 1.

These steady state solutions only exist in certain parameter
regimes.  For $\gamma<\alpha/(1+\alpha)$, instead of iterating to the
fixed point (\ref{alpha}), the density is fixed for $i>1$ at $\rho_i=
1-2\gamma$, with $\rho_1 = v/\gamma$ and
$v=\gamma(1-2\gamma)/(1-\gamma)$. The interpretation is that the rate
of release of particles at the growing end is no longer large enough
to control the input rate.  Thus the particle density reaches a
maximum value that extends from near the tip throughout the whole
lattice and the boundary condition (\ref{lbc}) is not satisfied.  This
is not a steady state solution for our model in the sense we have
defined and we describe this region as a ``jammed'' phase.  At the
transition to the jammed phase the bulk density jumps discontinuously
from $\rho= \alpha = \gamma/(1-\gamma)$ to $\rho = 1-2\gamma$.

\begin{figure}
\centerline{
\mbox{
\includegraphics[width=2.5in,angle=270]{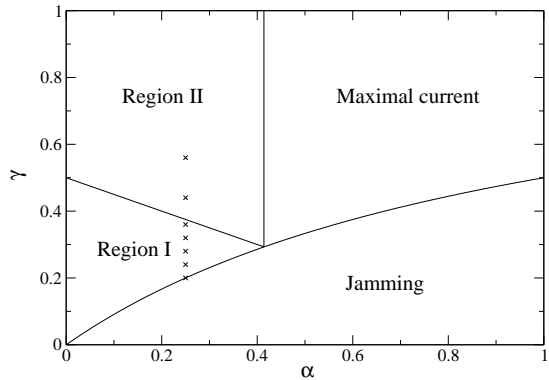}
}
}
\caption{Phase diagram given by the simple mean-field theory presented
  in the text, for the model with $\beta=0$. Phases are discussed in
  the text.  Sample points for MC simulations of Fig.~\ref{MC} are
  marked with x's.}
\label{PD} 
\end{figure}

We see from (\ref{alpha}) that the maximum value of $\alpha$ is
$\alpha_c=\sqrt{2}-1$, which is obtained when $v= 3 -2\sqrt{2}$.  For
$\alpha >\alpha_c$, Eq. \ref{rhoi} no longer has real fixed points,
and again we do not satisfy the boundary condition (\ref{lbc}).  We
may understand the region bounded by $\alpha>\sqrt{2}-1$ and
$\gamma>\alpha/(1+\alpha)$ as a maximal current phase, where the
particles have reached a maximum flow rate through the system which is
no longer limited by the input and growth rates.  In this case, the
density profile decays algebraically from the boundary sites $1$ and
$N$ to a bulk density $\rho= \sqrt{2}-1$ and does not constitute a
steady state in our sense since the densities evolve as the system
grows.

\begin{figure}
\centerline{
\mbox{
\includegraphics[width=2.5in,angle=270]{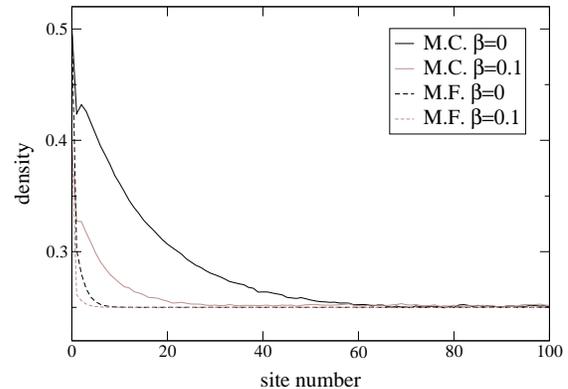}
}
}
\caption{Mean-field density profiles compared to MC density profiles
for $\beta=0$ and $\beta>0$. $\alpha=0.25$, $\gamma=0.3$}
\label{compare} 
\end{figure}

We summarise the results of this mean-field theory with a phase diagram
in Fig.~\ref{PD}.  Regions I and II correspond to the MC profiles
observed for high $\gamma$ (Fig.~\ref{MC}), and the jammed region
corresponds to the profiles observed for low $\gamma$.  Simulations
carried out over the whole parameter space revealed that the
qualitative behaviour of the mean-field theory is correct, however,
the transitions between different profile types do not occur exactly
at the predicted mean-field boundaries. In order to compare in more
detail the mean field and MC results we plot in Fig.~\ref{compare}
mean-field and simulation profiles in region I.  The decay length at
the tip is significantly higher in the simulation, by a factor of
about 10.  The differences between the mean-field theory presented
here, and simulations can be attributed to density fluctuations and
correlations in the system which are ignored in the mean-field theory.
In particular the occupations of sites 1 and 2 are strongly correlated
since a growth event vacates both of these sites simultaneously.  An
improved mean-field theory which takes into account the correlation
between sites 1 and 2 predicts a phase diagram with essentially the
same phases, but some modified phase boundaries \cite{SE}.

Simulation and mean-field results for non-zero $\beta$,
Fig.~\ref{compare}, show that $\beta$ does not affect the qualitative
profile shape.  A detailed analysis of the phase structure with
$\beta\neq 0$ will be given elsewhere \cite{SE}.

We take particular interest in the steady state phase with a
positive density gradient at the tip, and speculate whether the
emergence of this high density region may be associated with the
vesicle accumulation within the Spitzenko\"rper in the hyphal tip
region of a real fungus.  For representative parameter values
$\alpha=0.25$ and $\gamma=0.24$, MC simulations predict a high density
region of $\sim 240$~nm, correlating (within an order of magnitude) with
the length scale of the Spitzenko\"rper in {\em N. crassa} of
$\lesssim 2 \mu$m \cite{26,caution}.

Note that, in contrast to the vesicle supply centre model
\cite{Garcia,TKM}, a growth velocity arises naturally for our model. At
least within mean-field theory, this velocity is determined in a
simple way by the rate $\alpha$ at which material is fed into the
system far away from the tip and the ratio $\beta/\gamma$ representing
the efficiency with which the cargoes fuse with the tip.  It would be
of interest to test this prediction experimentally, for example by
live-cell imaging the dynamics of motor proteins or cargo fused with
the green flourescent protein \cite{GFP}.  In this way the rate of
delivery of motor proteins (or cargo) to the hyphal tip could be analysed
in relation to the rate of hyphal extension (which can be controlled,
for example, by changing the temperature of incubation).
Finally we mention that the phase diagram Fig.~\ref{PD} is related to
that of the open boundary TASEP \cite{TASEP}; we explore this
correspondence in a future publication \cite{SE}.

We thank Graham Wright for many helpful discussions on fungal biology.

KS is funded by the EPSRC.


\begin{thebibliography}{99999}
\bibitem{caution} We should treat this comparison with caution: a real
hypha tapers at the tip; this has to be taken into account when
comparing observed vesicle concentrations with site occupancy in our
strictly 1D model.
\end{thebibliography}
\end{document}